\documentclass[sigconf]{acmart}
\AtBeginDocument{%
  }

\setcopyright{acmlicensed}

\renewcommand\footnotetextcopyrightpermission[1]{}

\copyrightyear{2018}
\acmYear{2018}
\acmDOI{XXXXXXX.XXXXXXX}

\acmConference[Conference acronym 'XX]{Make sure to enter the correct
  conference title from your rights confirmation email}{June 03--05,
  2018}{Woodstock, NY}
\acmISBN{978-1-4503-XXXX-X/2018/06}

\usepackage{subcaption}
\usepackage{multirow}
\usepackage{enumitem}
\usepackage{graphicx}
\usepackage{balance}
\usepackage{amsmath}
\usepackage{booktabs}
\usepackage{amssymb}
\usepackage[capitalize]{cleveref}
\usepackage{pifont}

\usepackage{amsmath}
\usepackage{algorithm}
\usepackage{algpseudocode}
\usepackage{amsfonts}

\usepackage{listings}
\usepackage{url}
\usepackage{hyperref}

\usepackage{colortbl}  
\usepackage{xcolor}
\usepackage{array}   
\usepackage{marvosym} 

\newcommand{\eg}{\textit{e}.\textit{g}.}

\definecolor{linecolor}{gray}{.91} 
\definecolor{linecolor2}{gray}{.95} 
\definecolor{linecolor1}{gray}{.97} 

\usepackage{color, xspace}
\usepackage[normalem]{ulem}

\lstdefinestyle{custom_style}{
    basicstyle=\ttfamily\small,      
    breaklines=true,                  
    frame=single,                     
    xleftmargin=1pt,                
    xrightmargin=1pt,               
    keywordstyle=\color{blue},       
    commentstyle=\color{gray},       
    stringstyle=\color{red},         
}

\begin{document}


\title{HiD-VAE: Interpretable Generative Recommendation via Hierarchical and Disentangled Semantic IDs}

\author{Dengzhao Fang}
\affiliation{%
  \institution{Jilin University}
    \city{Changchun}
  \country{China}
}
\email{fangdz25@mails.jlu.edu.cn}

\author{Jingtong Gao}
\affiliation{%
  \institution{City University of Hong Kong}
    \city{Hong Kong}
  \country{China}
}
\email{jt.g@my.cityu.edu.hk}

\author{Chengcheng Zhu}
\affiliation{%
  \institution{Nanjing University}
    \city{Nanjing}
  \country{China}
}
\email{602025320026@smail.nju.edu.cn}

\author{Yu Li}
\affiliation{%
  \institution{Jilin University}
    \city{Changchun}
  \country{China}
}
\email{liyu90@jlu.edu.cn}

\author{Xiangyu Zhao}
\affiliation{%
  \institution{City University of Hong Kong}
    \city{Hong Kong}
  \country{China}
}
\email{xianzhao@cityu.edu.hk}

\author{Yi Chang}
\affiliation{%
  \institution{Jilin University}
    \city{Changchun}
  \country{China}
}
\email{yichang@jlu.edu.cn}


\renewcommand{\shortauthors}{Trovato et al.}

\begin{abstract}

Recommender systems are indispensable for helping users navigate the immense item catalogs of modern online platforms. Recently, generative recommendation has emerged as a promising paradigm, unifying the conventional retrieve-and-rank pipeline into an end-to-end model capable of dynamic generation. However, existing generative methods are fundamentally constrained by their unsupervised tokenization, which generates semantic IDs suffering from two critical flaws: (1) they are semantically flat and uninterpretable, lacking a coherent hierarchy, and (2) they are prone to representation entanglement (i.e., ``ID collisions''), which harms recommendation accuracy and diversity. To overcome these limitations, we propose HiD-VAE, a novel framework that learns hierarchically disentangled item representations through two core innovations. First, HiD-VAE pioneers a hierarchically-supervised quantization process that aligns discrete codes with multi-level item tags, yielding more uniform and disentangled IDs. Crucially, the trained codebooks can predict hierarchical tags, providing a traceable and interpretable semantic path for each recommendation. Second, to combat representation entanglement, HiD-VAE incorporates a novel uniqueness loss that directly penalizes latent space overlap. This mechanism not only resolves the critical ID collision problem but also promotes recommendation diversity by ensuring a more comprehensive utilization of the item representation space. These high-quality, disentangled IDs provide a powerful foundation for downstream generative models. Extensive experiments on three public benchmarks validate HiD-VAE's superior performance against state-of-the-art methods.

\end{abstract}


\maketitle

\section{Introduction}
\label{sec:introduction}
\begin{figure}[!t]
    \begin{center}
    \includegraphics[width=\columnwidth]{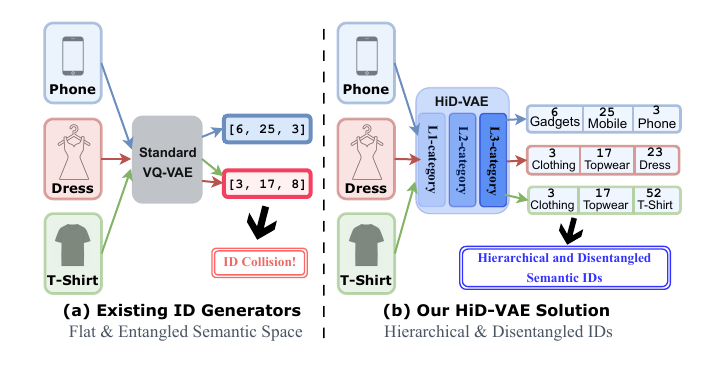}
    \vskip -0.1in
    \caption{\textbf{Comparison of Semantic ID Generation.} 
    \textbf{(a)}: Conventional methods learn a flat semantic space, leading to ``ID collisions'' where distinct items (e.g., a ``Dress'' and ``T-Shirt'') are mapped to the same ID. 
    \textbf{(b)}: Our HiD-VAE learns a hierarchical, disentangled space. It generates specific, layered semantic IDs, a structure that not only prevents collisions but also enhances recommendation interpretability by providing a clear category path (e.g., Clothing $\rightarrow$ Topwear $\rightarrow$ Dress).
    }
    \label{fig:intro}
    \end{center}
    \vskip -0.1in
\end{figure}

Recommender systems are essential for navigating information overload on modern digital platforms. Among these, sequential recommendation has become a cornerstone for capturing the dynamic nature of user preferences \cite{10.1145/3539618.3591717, 10.1145/3637871,chen2021modeling}. The field has been shaped by a progression of deep learning architectures; early pioneering work utilized recurrent neural networks in models like GRU4Rec~\cite{hidasi2015session} and convolutional neural networks in models like Caser~\cite{tang2018personalized}. More recently, Transformer-based models\cite{vaswani2017attention}, such as SASRec~\cite{kang2018self} and BERT4Rec~\cite{sun2019bert4rec}, have excelled at modeling complex dependencies.


While these traditional models excel at scoring candidates within a representation matching framework, the field has recently witnessed a paradigm shift towards generative recommendation~\cite{rajput2023recommender,zheng2024adapting,wang2024eager,deng2025onerec,li2025semantic}. This shift is driven by the paradigm's potential to unify the conventional retrieve-and-rank pipeline into an end-to-end model that autoregressively generates item identifiers. Pioneering works like TIGER~\cite{rajput2023recommender} introduced ``Semantic IDs'', the identifiers that distill rich semantic features into numerical codes to represent items. However, the efficacy of this entire generative pipeline is critically dependent on the quality of the generated semantic IDs~\cite{hou2025generative,wang2023generative,deldjoo2024review}, which has surfaced some fundamental challenges.

The generative recommendation paradigm typically operates as a two-stage process: first, a codebook is learned to quantize the content features of each item into a semantic ID. Second, a sequence model is trained to autoregressively predict the semantic ID of the next item based on user history~\cite{rajput2023recommender,wang2024eager,zheng2024adapting}. Recent advancements have focused on enhancing one of these two stages: some, like LETTER~\cite{wang2024learnable}, improve the quantization stage by injecting collaborative signals, while others, like LC-Rec~\cite{zheng2024adapting}, refine the sequence model training through sophisticated alignment tasks. The quantization stage is paramount, as the quality of the learned IDs forms the foundation for the entire generative pipeline \cite{jia2025principles,hua2023index,lin2025order,li2024survey}. Despite its advances, the predominant approach, relying on unsupervised vector quantization (VQ)~\cite{gray1984vector} with techniques like RQ-VAE~\cite{zeghidour2022soundstream} or PQ~\cite{hou2023learning}, is hampered by two fundamental challenges. First, it produces flat, uninterpretable semantic IDs~\cite{liu2021interpretable,deldjoo2024review}, where any learned hierarchy is merely an implicit byproduct rather than an explicitly supervised structure, rendering the models uncontrollable ``black boxes''. Second, it suffers from severe representation entanglement \cite{wang2024disentangled,ma2019learning,du2024disentangled,wang2022disentangled}, which leads to ``ID collisions''~\cite{rajput2023recommender,wu2025graphhash}—a critical flaw where distinct items are erroneously mapped to the same identifier. This entanglement, often addressed with ineffective post-hoc fixes~\cite{rajput2023recommender}, undermines recommendation accuracy and diversity. As illustrated in Figure~\ref{fig:intro}~(a), this dual bottleneck in tokenization limits the full potential of current generative models.

To address these dual challenges, we introduce HiD-VAE, a novel framework that learns interpretable and disentangled item representations, as depicted in Figure~\ref{fig:intro} (b). To tackle the challenge of ID collisions and interpretability, HiD-VAE first pioneers a hierarchically-supervised quantization process using hierarchical tags. This is enforced through a tag alignment loss and a tag prediction loss, which explicitly guide each VAE layer to capture a specific level of category semantics. Concurrently, to combat representation entanglement and prevent ID collisions at the source, we design a novel uniqueness loss that directly penalizes the latent representation overlap between distinct items. Furthermore, to ensure our framework's applicability to real-world datasets that often lack ground-truth labels, we further introduce an effective LLM-based approach for automatically generating hierarchical tags. We summarize our major contributions as follows:

\begin{itemize}[leftmargin=*]
    \item We propose HiD-VAE, a novel framework that learns hierarchically structured and disentangled representations specifically designed for the demands of generative recommendation.

    \item We introduce two core technical innovations to address key limitations in existing models: a hierarchically-supervised process, driven by tag alignment and prediction losses, to ensure an interpretable semantic hierarchy; and a uniqueness loss to enforce representation disentanglement and prevent ID collisions.

    \item We validate our framework through extensive experiments on three public datasets, demonstrating state-of-the-art performance in recommendation accuracy and interpretability. For datasets lacking explicit categorical structures (\eg, KuaiRand), we showcase our method's broad applicability by employing a practical, LLM-based strategy to generate high-quality hierarchical tags.

\end{itemize}

\section{Related Work}
\label{sec:related}

In line with our research focus, we briefly outline the research trajectory in the field of sequential recommendation, ranging from traditional sequential  models and transformer-based models to generative models, along with their most representative works. 
Due to space constraints, the detailed discussion of related work can be found in Appendix~\ref{sec:appendix_related}.


\section{Methodology}
\label{sec:method}

\begin{figure*}[t!]
  \centering 
  \includegraphics[width=0.86\linewidth]{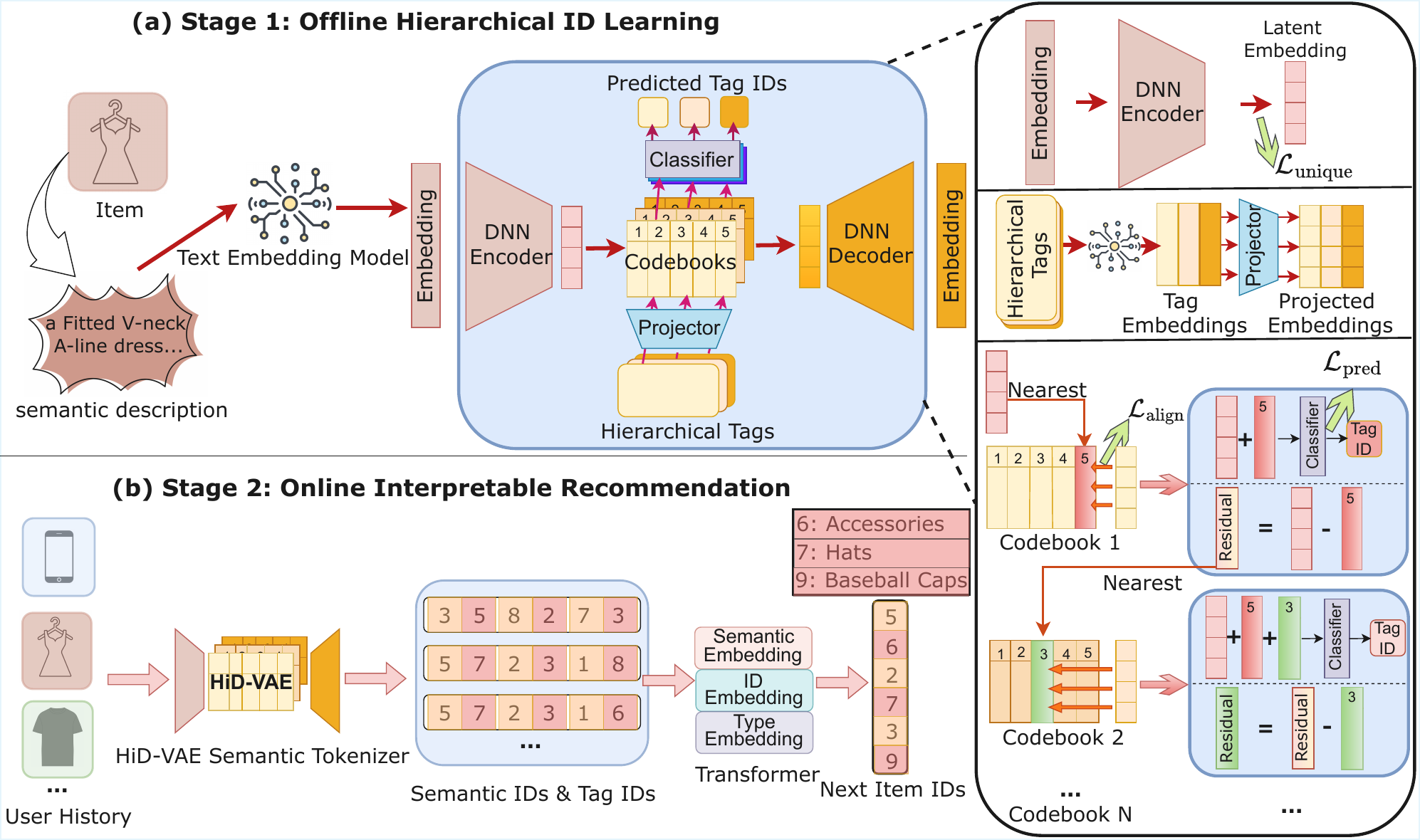}
  \vspace{-3mm}
  \caption{{\bf The HiD-VAE framework.} HiD-VAE first learns hierarchical semantic IDs and disentangled tag IDs via a supervised VAE (a), which a Transformer then uses for interpretable sequential recommendation (b).}
  \label{fig:framework}
  \vspace{-2mm}
\end{figure*}

This section details our proposed HiD-VAE framework. We first formulate the generative sequential recommendation task, then present our two-stage model and elaborate on its architecture.

\subsection{Preliminary}

\noindent\textbf{Variational Autoencoder.} A Variational Autoencoder (VAE)~\cite{kingma2013auto} is a generative model comprising an encoder $E(\cdot)$ and a decoder $D(\cdot)$. The encoder maps an input feature vector $\boldsymbol{x} \in \mathbb{R}^{d_{\text{in}}}$ to a continuous latent representation $\boldsymbol{z} = E(\boldsymbol{x}) \in \mathbb{R}^d$, from which the decoder reconstructs the input as $\hat{\boldsymbol{x}} = D(\boldsymbol{z})$. The model is optimized by minimizing reconstruction loss alongside a KL divergence term that regularizes the latent space.

\noindent\textbf{Vector Quantization (VQ)~\cite{gray1984vector}.} To introduce discreteness into the latent space, the VAE framework can be extended with Vector Quantization. VQ maps the continuous latent vector $\boldsymbol{z}$ to its nearest neighbor in a finite, learnable codebook $\mathcal{C} = \{\boldsymbol{c}_k\}_{k=1}^K$, where each codeword $\boldsymbol{c}_k \in \mathbb{R}^d$. This process, often referred to as VQ-VAE~\cite{van2017neural}, uses the following quantization function:
\begin{equation}
q(\boldsymbol{z}) = \boldsymbol{c}_{k^*} \quad \text{where} \quad k^* = \arg\min_j \|\boldsymbol{z} - \boldsymbol{c}_j\|_2
\end{equation}


\noindent\textbf{Residual-Quantized VAE (RQ-VAE)~\cite{zeghidour2022soundstream}.} RQ-VAEs further enhance this approach by employing a cascade of $L$ quantizers. Instead of quantizing the latent vector in one step, each subsequent quantizer operates on the residual error from the preceding stage, allowing for a more efficient and fine-grained discrete representation. We define the cumulative quantized embedding up to layer $l$ as $\boldsymbol{z}_q^{(l)} = \sum_{j=1}^{l} \boldsymbol{e}^{(j)}$, where $\boldsymbol{e}^{(j)}$ is the selected codeword from the $j$-th quantizer. In our work, the cascaded architecture from RQ-VAE is adapted to implement our hierarchical item tokenizer. We then introduce explicit hierarchical supervision and a novel disentanglement mechanism to address the inherent limitations of this unsupervised framework in the recommendation context.

\subsection{Problem Formulation}

Let $\mathcal{U}$ be the set of users and $\mathcal{I}$ be the set of items. Each user $u \in \mathcal{U}$ has a chronological interaction sequence $\mathcal{S}_u = (i_1, i_2, \ldots, i_T)$, where $T = |\mathcal{S}_u|$ is the sequence length and $i_t \in \mathcal{I}$. The goal of sequential recommendation is to predict the next item $i_{T+1}$ that user $u$ is most likely to interact with.

In the generative recommendation approach, we reframe this as generating a unique ID for the target item. Each item $i$ is represented by a structured, hierarchical semantic ID $\boldsymbol{y}_i = (y_i^{(1)}, y_i^{(2)},\ldots, y_i^{(L)})$, where $L$ is the number of levels (ID length), and each $y_i^{(l)}$ is an index in a level-specific codebook $\mathcal{C}^{(l)}$ of size $K_l$. These IDs are learned to match human-understandable item categories. The task then becomes an autoregressive prediction problem:

\begin{equation}
p(\boldsymbol{y}_{t+1} | \boldsymbol{y}_1, \ldots, \boldsymbol{y}_t) = \prod_{l=1}^{L} p(y_{t+1}^{(l)} | \boldsymbol{y}_1, \ldots, \boldsymbol{y}_t, y_{t+1}^{(1)}, \ldots, y_{t+1}^{(l-1)})
\end{equation}

\subsection{Framework Overview}

The HiD-VAE framework cleanly separates representation learning from sequential modeling across two stages.

\noindent\textbf{Stage 1: Offline Hierarchical ID Learning.} As illustrated in Figure \ref{fig:framework}(a), we first train our hierarchical and disentangled VAE (HiD-VAE) on the entire dataset. For each item defined by content features $\boldsymbol{x}$ (text embeddings), HiD-VAE learns a hierarchical semantic ID $\boldsymbol{y}$. This is accomplished by optimizing a composite objective function that incorporates standard VAE losses with our novel hierarchically-supervised and uniqueness losses. The output is a frozen, high-quality item tokenizer capable of converting any item into its unique, interpretable, and disentangled semantic ID.

\noindent\textbf{Stage 2: Online Interpretable Recommendation.} As illustrated in Figure \ref{fig:framework}(b), the pre-trained and frozen HiD-VAE serves as the item tokenizer. For each user's interaction history, every item is mapped to its corresponding semantic ID sequence. A Transformer-based sequential model, equipped with hierarchy-aware semantic embeddings to preserve the structured semantics from Stage 1, is then trained to autoregressively predict the semantic ID of the next item. During inference, constrained decoding ensures that generated IDs correspond to valid items. This two-stage design allows each component to be specialized for its specific task.

\subsection{Hierarchical Tag Generation}\label{sec:Hierarchical_Tag_Generation}



Hierarchical categorization and labeling of items are essential for enhancing recommendation systems by minimizing the likelihood of ID collisions and fostering improved interpretability. 
Many real-world datasets are often devoid of hierarchical tags, making manual annotation both labor-intensive and costly, with inherent inconsistencies. 
Current unsupervised clustering methods tend to yield results that are difficult to interpret, while general LLMs face challenges related to hallucinations~\cite{huang2025survey}. 
To overcome these limitations, we introduce a two-stage constrained workflow that utilizes LLMs for hierarchical tag generation. 
This innovative approach reformulates the task into a “retrieval-then-classification” pipeline, facilitating the generation of reliable hierarchical tags without the need for manual intervention, thereby reducing potential risks. 

\noindent\textbf{Candidate Tag Retrieval.} The first stage aims to narrow down the vast universe of possible tags to a small and relevant candidate set for each hierarchical level. We begin by constructing a tag pool for each level $l$, denoted as $\mathcal{T}^{(l)}$, from existing labels and manual annotations.
Specifically, for an item with content text $\boldsymbol{x}_{\text{text}}$, we first compute its semantic embedding $\boldsymbol{v} = \mathcal{E}_{\text{sent}}(\boldsymbol{x}_{\text{text}})$ using a pre-trained sentence encoder. Then, for each level $l$, we retrieve the top-K most semantically similar tags by performing a nearest-neighbor search against the pre-computed tag embeddings in the corresponding index. The candidate set $\mathcal{C}_{\text{cand}}^{(l)}$ is defined as:
\begin{equation}
\mathcal{C}_{\text{cand}}^{(l)} = \text{Top-K}_{t \in \mathcal{T}^{(l)}} \left( \text{sim}(\boldsymbol{v}, \mathcal{E}_{\text{sent}}(t)) \right)
\end{equation}

\noindent\textbf{LLM-based Tag Classification.} With a small candidate set, we can now leverage the capabilities of an LLM without exceeding its context window or risking hallucination. We reformulate the task as a classification problem, prompting the LLM to select the most suitable tag from the candidate list. The prompt is constructed to include the item's content $\boldsymbol{x}_{\text{text}}$, any previously offered higher-level tags $\{t^{(j)}\}_{j=1}^{l-1}$, and the candidate set $\mathcal{C}_{\text{cand}}^{(l)}$. The LLM's task is to predict the most probable tag $t^{(l)*}$ for the current level $l$:
\begin{equation}
t^{(l)*} = \underset{t \in \mathcal{C}_{\text{cand}}^{(l)}}{\arg\max} \, P_{\text{LLM}}\left(t \mid \boldsymbol{x}_{\text{text}}, \{t^{(j)}\}_{j=1}^{l-1}, \mathcal{C}_{\text{cand}}^{(l)}\right)
\end{equation}
See Appendix~\ref{prompt} for details on LLM selection and prompt design. This ``retrieval-then-classification'' approach ensures that the generated tags are always valid and selected based on the deep contextual understanding of the LLM, providing a robust solution for extending our framework to a wider range of datasets.





\subsection{Hierarchical Representation Learning}
\label{sec:hierarchical_supervision}




Existing unsupervised VAE-based tokenizers frequently exhibit codebook collapse, resulting in significant ID collisions where different items share identical representations, which adversely affects recommendation accuracy and diversity. 
This issue arises from their flat, unstructured latent spaces that inadequately capture hierarchical semantics, rendering IDs ambiguous and susceptible to entanglement. 
To counteract this limitation, HiD-VAE employs hierarchically-supervised quantization, which aligns the discrete codes from each RQ-VAE layer with multi-level item tags. 
This approach establishes a more structured representation space, effectively reducing collapse and improving semantic fidelity.

Given an item with feature vector $\boldsymbol{x}$ and ground-truth hierarchical category tags with indices $\{c^{(l)}\}_{l=1}^L$ and embeddings $\{\boldsymbol{t}^{(l)}\}_{l=1}^L$, the process passes the item features through encoder $E(\cdot)$ to produce an initial latent representation $\boldsymbol{z}_0 = E(\boldsymbol{x})$.

The RQ-VAE then initiates a layered quantization process. At each layer $l \in \{1, \ldots, L\}$, quantizer $q_l$ takes the residual from the previous layer $\boldsymbol{r}_{l-1}$ (with $\boldsymbol{r}_0 = \boldsymbol{z}_0$) and identifies the closest codeword $\boldsymbol{e}^{(l)} = q_l(\boldsymbol{r}_{l-1})$ from codebook $\mathcal{C}^{(l)}$. The residual for the next layer is computed as $\boldsymbol{r}_l = \boldsymbol{r}_{l-1} - \boldsymbol{e}^{(l)}$.

\noindent\textbf{Tag Alignment Loss.} To ensure that the learned codebook at layer $l$ captures the semantics of the $l$-th level of the category hierarchy, we introduce a contrastive tag alignment loss. We project the ground-truth tag embedding $\boldsymbol{t}^{(l)}$ into the item's latent space using a layer-specific projector $P_l(\cdot)$. The loss pulls the cumulative quantized embedding $\boldsymbol{z}_q^{(l)}$ towards its corresponding projected tag embedding $P_l(\boldsymbol{t}^{(l)})$ while pushing it away from other tag embeddings in the same mini-batch:

\begin{equation}
\mathcal{L}_{\text{align}}^{(l)} = -\log \frac{\exp(\text{sim}(\boldsymbol{z}_q^{(l)}, P_l(\boldsymbol{t}^{(l)})) / \tau)}{\sum_{j=1}^{B} \exp(\text{sim}(\boldsymbol{z}_q^{(l)}, P_l(\boldsymbol{t}_j^{(l)})) / \tau)},
\end{equation}
where $\text{sim}(\cdot, \cdot)$ denotes cosine similarity, $\tau$ is a temperature hyperparameter, and $B$ is the batch size.

\noindent\textbf{Tag Prediction Loss.} Hierarchical tags exhibit varying semantic depths and numbers of categories, with finer-grained tags at deeper layers encompassing more classes and requiring higher prediction difficulty due to increased specificity. To accommodate this, each layer employs a tailored classifier $C_l$ whose structure scales with the layer depth: specifically, deeper layers utilize larger hidden dimensions and progressively higher dropout~\cite{srivastava2014dropout} rates to handle the richer cumulative quantized embedding $\boldsymbol{z}_q^{(l)}$ as input, which concatenates embeddings from all preceding layers and thus grows in dimensionality and informational complexity. This design ensures that classifiers for deeper layers have greater parameter capacity to capture nuanced semantics. The loss for each layer is computed as:

\begin{equation}
\mathcal{L}_{\text{pred}}^{(l)} = \text{CrossEntropy}(C_l(\boldsymbol{z}_q^{(l)}), c^{(l)}),
\end{equation}
where optional extensions like focal loss (with $\gamma=2.0$) can be applied for imbalanced classes, as implemented in our framework.

This dual supervision ensures the learned semantic IDs are structured and mapped to an interpretable hierarchy.

\subsection{Disentanglement via Uniqueness Loss}
\label{sec:disentanglement}

A critical challenge in generative recommendation is ``ID collision'', where distinct items are mapped to identical discrete ID sequences. To combat this issue, we introduce a uniqueness loss that operates on the continuous, pre-quantization latent vectors.

The loss penalizes representation overlap between pairs of distinct items that are assigned identical semantic ID sequences within a training batch. Let $\boldsymbol{x}_i$ and $\boldsymbol{x}_j$ be two different items in a batch, with their initial latent representations being $\boldsymbol{z}_{0,i}$ and $\boldsymbol{z}_{0,j}$ respectively. If their full semantic ID sequences collide (i.e., $\boldsymbol{y}_i = \boldsymbol{y}_j$), we apply a margin-based penalty:

\begin{equation}
\mathcal{L}_{\text{unique}} = \frac{1}{|\mathcal{P}|} \sum_{(i,j) \in \mathcal{P}} \max\left(0, \frac{\boldsymbol{z}_{0,i} \cdot \boldsymbol{z}_{0,j}}{\|\boldsymbol{z}_{0,i}\|_2 \|\boldsymbol{z}_{0,j}\|_2} - m\right),
\end{equation}
where $\mathcal{P} = \{(i,j) | i \neq j, \boldsymbol{y}_i = \boldsymbol{y}_j\}$ is the set of all distinct item pairs with colliding IDs within the batch, and $m$ is a margin hyperparameter. This loss directly encourages an injective mapping from items to IDs, mitigating representation entanglement.

\subsection{Interpretable Generative Recommendation}\label{sec:Interpretable_Generative-Recommendation}

With the high-quality IDs generated in Stage 1, Stage 2 leverages these for sequential recommendation, where a generative model predicts the next item's ID based on a user's interaction history. However, modeling these structured, multi-level IDs for autoregressive generation poses challenges that a standard Transformer cannot adequately address: it may treat the semantic ID sequence (e.g., $\boldsymbol{y}_i = (y_i^{(1)}, \ldots, y_i^{(L)})$) as a flat token stream, losing the structured semantics encoded by HiD-VAE, while unconstrained generation risks producing invalid ID combinations that do not correspond to real items, undermining practical applicability \cite{rajput2023recommender}.

To address these issues, we propose a tailored Transformer-based autoregressive model with two key innovations:
\begin{itemize}[leftmargin=*]    
        \item \textbf{Hierarchy-Aware Semantic Embeddings:} To preserve the structured semantics of our IDs and enhance interpretability, we design a custom embedding layer. Each token in an item's semantic ID is first mapped to its corresponding tag text, which is then encoded into a semantic vector using a pre-trained embedding model. These semantic vectors are concatenated with learnable ID embeddings and type embeddings specific to each hierarchical level ($l \in {1, \ldots, L}$). This approach enriches the feature representation by integrating explicit semantic information, enabling the model to capture the coarse-to-fine semantic path encoded in the ID while enhancing interpretability.
        
        \item \textbf{Constrained Decoding for Validity:} To ensure generated IDs correspond to real items, we implement a constrained decoding strategy during inference. We pre-compute and store all valid semantic ID prefixes in an efficient data structure. During token-by-token generation, the model's output vocabulary is dynamically masked to allow only tokens that form valid, existing prefixes. This pruning mechanism guarantees that the generated output always corresponds to an actual item in the inventory.
\end{itemize}

\subsection{Optimization}

The framework is trained in two distinct stages, each with a tailored objective function.

\noindent\textbf{Stage 1: HiD-VAE Training.} The HiD-VAE is trained end-to-end by minimizing a composite loss function:


\begin{equation}
    \begin{aligned}
        \mathcal{L}_{\text{HiD-VAE}} = & \mathcal{L}_{\text{recon}} + \beta_{\text{commit}} \mathcal{L}_{\text{commit}} \\
& + \beta_{\text{sup}} \sum_{l=1}^{L} \big(\mathcal{L}_{\text{align}}^{(l)} + \mathcal{L}_{\text{pred}}^{(l)}\big) + \beta_{\text{unique}} \mathcal{L}_{\text{unique}}
    \end{aligned}
\end{equation}
where $\mathcal{L}_{\text{recon}}$ is the reconstruction loss between the input $\boldsymbol{x}$ and the decoder output $\hat{\boldsymbol{x}}$, and $\mathcal{L}_{\text{commit}}$ is the vector quantization commitment loss, which regularizes the encoder's output space~\cite{kingma2013auto, van2017neural}, and $\beta_{\text{commit}}$, $\beta_{\text{sup}}$, $\beta_{\text{unique}}$ are hyperparameters that balance the loss components. Details for these standard loss $\mathcal{L}_{\text{recon}}$ and $\mathcal{L}_{\text{commit}}$ components are provided in Appendix~\ref{autoencoder} for completeness.

\noindent\textbf{Stage 2: Recommender Training.} With HiD-VAE parameters frozen, the Transformer-based recommender is trained using next-token prediction with cross-entropy loss. For a user history $\mathcal{S}_u$ with ID sequences $(\boldsymbol{y}_1, \ldots, \boldsymbol{y}_T)$, the objective maximizes the likelihood of the next item's ID sequence:

\begin{equation}
\mathcal{L}_{\text{rec}} = - \sum_{u \in \mathcal{U}} \sum_{t=1}^{|\mathcal{S}_u|-1} \log p(\boldsymbol{y}_{t+1} | \boldsymbol{y}_1, \ldots, \boldsymbol{y}_t)
\end{equation}

To enhance training, each token in the semantic ID is mapped to its corresponding tag text and encoded into a semantic vector using a pre-trained embedding model, enriching the representation with explicit semantic information.

This two-stage approach first establishes an interpretable and disentangled representation space, then leverages it to model sequential user behavior effectively, see Appendix~\ref{algorithm} for details.


\section{Experiments}
\label{sec:exp}
In this section, we conduct extensive experiments to rigorously evaluate our proposed HiD-VAE framework. Our evaluation is designed to answer the following key research questions:
\begin{itemize}[leftmargin=*]
    \item \textbf{RQ1:} How does HiD-VAE perform compared to state-of-the-art traditional sequential and generative recommendation baselines on multiple public benchmarks?
    \item \textbf{RQ2:} What are the individual contributions of our core technical innovations: the hierarchically-supervised process and the disentanglement via uniqueness loss?
    \item \textbf{RQ3:} Does HiD-VAE successfully learn an interpretable and semantically structured latent space? 
\end{itemize}

\subsection{Experimental Settings}

\noindent\textbf{Datasets.}
We evaluate our model on three widely-used public benchmarks to ensure a comprehensive assessment of its capabilities across different domains and data characteristics. Following standard practice~\cite{kang2018self,sun2019bert4rec}, we adopt the 5-core setting, where all users and items with fewer than five interactions are filtered out.

\begin{itemize}[leftmargin=*]
    \item We evaluate our framework on two widely-used datasets derived from the Amazon Review Data project~\cite{he2016ups}\footnote{Available at: \url{https://jmcauley.ucsd.edu/data/amazon/}}.
    \begin{itemize}[leftmargin=1em, topsep=2pt, itemsep=1pt]
        \item \textbf{Beauty:} This dataset serves as a popular and relatively dense benchmark for recommendation research.
        \item \textbf{Sports and Outdoors:} In contrast, this dataset is larger and significantly sparser, allowing us to evaluate our model's robustness under more challenging data distributions.
    \end{itemize}
    
    \item \textbf{KuaiRand-1K}~\cite{gao2022kuairand}\footnote{Available at: \url{https://kuairand.com/}} This is a large-scale public dataset from the Kuaishou short video platform, containing user interactions with rich side information. It represents a distinct domain and serves to test the generalizability of our approach.

\end{itemize}

The detailed statistics of these datasets can be found in Appendix~\ref{app:dataset_stats}. 
For datasets like KuaiRand, where structured hierarchies are not natively provided, we use an LLM-based pre-processing pipeline. We prompt a large language model with the item's title and raw category string to generate a clean, consistent $L$-level category hierarchy. This ensures every item has a complete category path for supervision. See Appendix~\ref{prompt} for more details.

\noindent\textbf{Baselines.} We conduct a comprehensive comparison of HiD-VAE against a wide spectrum of state-of-the-art models, which we group into three distinct categories. 1) Traditional Sequential Models: GRU4Rec~\cite{hidasi2015session}, Caser~\cite{tang2018personalized}, HGN~\cite{ma2019hierarchical}, and NextItNet~\cite{yuan2019simple}; 2) Transformer-based Models: SASRec~\cite{kang2018self}, and BERT4Rec~\cite{sun2019bert4rec}; 3) Generative Recommendation Models: TIGER~\cite{rajput2023recommender}, LC-Rec~\cite{zheng2024adapting}, and VQ-Rec~\cite{hou2023learning}. 
See Appendix~\ref{baselines} for more details on baselines. 


\begin{table*}[t]
\centering
\caption{Overall performance comparison on three benchmark datasets. Metrics are abbreviated: Recall (R) and NDCG (N). The best results are in \textbf{bold}, and the second-best are \underline{underlined}. `Improv.' denotes the relative improvement of HiD-VAE over the strongest baseline. All improvements are statistically significant ($p < 0.01$).}
\vspace{-3mm}
\label{tab:main_results}
\resizebox{\textwidth}{!}{%
\begin{tabular}{clccccccccccc}
\toprule
\multirow{2}{*}{Dataset} & \multirow{2}{*}{Metric} & \multicolumn{4}{c}{\emph{Traditional}} & \multicolumn{2}{c}{\emph{Transformer-based}} & \multicolumn{4}{c}{\emph{Generative}} & \multirow{2}{*}{Improv.} \\
\cmidrule(lr){3-6} \cmidrule(lr){7-8} \cmidrule(lr){9-12}
& & GRU4Rec & Caser & HGN & NextItNet & SASRec & BERT4Rec & VQ-Rec & TIGER & LC-REC & HiD-VAE & \\
\midrule
\multirow{4}{*}{Beauty}
& R@5    & 0.0216 & 0.0093 & 0.0312 & 0.0143 & 0.0363 & 0.0116 & 0.0285 & 0.0312 & \underline{0.0402} & \textbf{0.0543} & +35.07\% \\
& R@10   & 0.0293 & 0.0146 & 0.0358 & 0.0221 & 0.0498 & 0.0174 & 0.0431 & 0.0457 & \underline{0.0563} & \textbf{0.0698} & +23.98\% \\
& N@5    & 0.0154 & 0.0058 & 0.0217 & 0.0090 & \underline{0.0269} & 0.0082 & 0.0182 & 0.0209 & 0.0257 & \textbf{0.0358} & +33.08\% \\
& N@10   & 0.0180 & 0.0075 & 0.0256 & 0.0115 & 0.0301 & 0.0100 & 0.0225 & 0.0253 & \underline{0.0366} & \textbf{0.0421} & +15.03\% \\
\midrule
\multirow{4}{*}{Sports}
& R@5    & 0.0097 & 0.0047 & 0.0162 & 0.0081 & 0.0202 & 0.0057 & 0.0291 & 0.0325 & \underline{0.0385} & \textbf{0.0435} & +12.99\% \\
& R@10   & 0.0150 & 0.0080 & 0.0235 & 0.0130 & 0.0290 & 0.0089 & 0.0415 & 0.0474 & \underline{0.0493} & \textbf{0.0632} & +28.19\% \\
& N@5    & 0.0065 & 0.0030 & 0.0111 & 0.0052 & 0.0118 & 0.0037 & 0.0199 & 0.0222 & \underline{0.0251} & \textbf{0.0332} & +32.27\% \\
& N@10   & 0.0082 & 0.0040 & 0.0134 & 0.0067 & 0.0146 & 0.0047 & 0.0238 & 0.0270 & \underline{0.0284} & \textbf{0.0397} & +39.79\% \\
\midrule
\multirow{4}{*}{KuaiRand}
& R@5    & 0.0298 & 0.0074 & 0.0297 & 0.0276 & 0.0332 & 0.0185 & 0.0513 & 0.0557 & \underline{0.0622} & \textbf{0.0668} & +7.40\% \\
& R@10   & 0.0383 & 0.0118 & 0.0354 & 0.0327 & 0.0405 & 0.0217 & 0.0589 & 0.0624 & \underline{0.0684} & \textbf{0.0785} & +14.77\% \\
& N@5    & 0.0217 & 0.0068 & 0.0169 & 0.0216 & 0.0338 & 0.0196 & 0.0354 & 0.0383 & \underline{0.0403} & \textbf{0.0479} & +18.86\% \\
& N@10   & 0.0245 & 0.0095 & 0.0219 & 0.0278 & 0.0372 & 0.0236 & 0.0412 & 0.0445 & \underline{0.0497} & \textbf{0.0586} & +17.91\% \\
\bottomrule
\end{tabular}
}
\end{table*}

\noindent\textbf{Evaluation Metrics.}
We adopt the standard leave-one-out~\cite{elisseeff2003leave} evaluation protocol. For each user's interaction history, the last item is held out for testing, the second-to-last item is used for validation, and the rest items are used for training. We report performance using two top-K ranking metrics: Recall@K and Normalized Discounted Cumulative Gain (NDCG)@K, with K set to {5, 10}.

\noindent\textbf{Implementation Details.}
Our framework is implemented in PyTorch with Hugging Face transformers and accelerate for mixed-precision (FP16) training on NVIDIA 4060 GPUs. 
For the representation learning stage, HiD-VAE uses a 3-layer MLP encoder and decoder with GELU activations, taking 768-dimensional SentenceTransformer embeddings as input. It employs $L=3$ quantization layers, each with codebook size $K=256$, initialized via K-Means on the first batch. Training uses AdamW with learning rate $3 \times 10^{-4}$ and batch size 128. Key loss hyperparameters are $\beta_{\text{commit}}=0.25$, $\beta_{\text{sup}}=1.0$, and $\beta_{\text{unique}}=2.0$. We apply Focal Loss for tag prediction, alignment temperature $\tau=0.07$, and uniqueness margin $m=0.9$. For the recommendation stage, the frozen HiD-VAE serves as item tokenizer with pre-computed semantic IDs. The sequential model is a 6-layer Transformer encoder-decoder with 8 attention heads and hidden dimension 512, trained using AdamW with learning rate $1 \times 10^{-4}$, batch size 256, and warmup schedule. Inference employs autoregressive generation with constrained decoding to prune invalid ID prefixes against the corpus cache. For sensitivity analysis on hyperparameters (e.g., codebook layers), see Appendix~\ref{hyperparam_sensitivity}.

For all traditional baseline models, we leverage the RecBole\footnote{\url{https://github.com/RUCAIBox/RecBole}}~\cite{zhao2021recbole} framework for implementation. For the LC-Rec baseline, to ensure a fair comparison in terms of model scale, we specifically utilize the T5-base variant as its backbone language model \cite{raffel2020exploring}.

\subsection{Overall Performance Comparison (RQ1)}

We present the comprehensive performance comparison of HiD-VAE against a suite of strong baselines in Table~\ref{tab:main_results}. The evaluation across three distinct datasets reveals several key insights:

\begin{itemize}[leftmargin=*]
    \item \textbf{HiD-VAE achieves substantial improvements across datasets.} HiD-VAE substantially outperforms all baselines on every dataset and metric,{The performance gains are particularly pronounced;} for instance, on the Beauty dataset, HiD-VAE achieves a remarkable 35.07\% relative improvement in Recall@5 and 33.08\% in NDCG@5 over the strongest baseline, LC-REC. Similar significant improvements are observed on the Sports (+32.27\% in NDCG@5) and KuaiRand (+18.86\% in NDCG@5) datasets. The superior performance is attributed to the  item IDs learned during Stage 1. By combining explicit hierarchical supervision with a novel disentanglement mechanism, HiD-VAE produces item representations that are not only semantically rich and interpretable but also uniquely distinct, providing a much stronger foundation for downstream generative recommender.

    \item \textbf{Structured representations enhance generative model efficacy.} While generative approaches represent the current frontier, our model surpasses all of them by a significant margin. Furthermore, a clear performance hierarchy emerges within the generative model family, underscoring the importance of the identifier's structure. VQ-Rec, which we adapted for this generative task using its semantic IDs, consistently underperforms TIGER and LC-Rec. This suggests that its non-hierarchical identifiers are less suited for autoregressive decoding, as the Transformer must predict a sequence of independent codes without the benefit of a coarse-to-fine semantic structure, likely leading to greater error propagation. 
    While TIGER and LC-Rec improve upon this with hierarchical codes based on RQ-VAE, they are still limited by their unsupervised nature, making them prone to semantic drift and ID collisions. This is where HiD-VAE excels. Our Hierarchically-Supervised (HS) process ensures each level of the ID aligns with a meaningful category, while the Disentanglement via Uniqueness Loss (DUL) actively minimizes collisions. This results in a more robust and disentangled representation space that is easier for the subsequent Transformer to model, leading to more precise and relevant recommendations.

    \item \textbf{Generative paradigms outperform discriminative methods in sequential recommendation.} A broader observation is the general superiority of generative models (HiD-VAE, LC-REC, TIGER) over traditional and Transformer-based sequential models that rely on discriminative scoring. This trend suggests that the paradigm of directly generating item identifiers, rather than scoring a pre-selected candidate set, is a more powerful approach. It sidesteps the potential disconnect between representation learning and the search/ranking process inherent in methods relying on Approximate Nearest Neighbor search. By directly modeling the probability distribution over the entire item universe (as represented by our discrete IDs), HiD-VAE can capture more complex and nuanced user preference patterns, confirming the advantages of our proposed ``learn-then-generate'' two-stage approach.

\end{itemize}


\subsection{Ablation Study (RQ2)}

To precisely isolate and quantify the contribution of each core component within our HiD-VAE framework, we conduct a granular ablation study. We design and evaluate the following model variants, systematically deactivating each key mechanism:

\begin{itemize}[leftmargin=*]
    \item \textbf{HiD-VAE (Full Model):} Our complete proposed model, which integrates the full Hierarchical Supervision (HS) mechanism, including both the Tag Alignment Loss and the Tag Prediction Loss, alongside the Disentanglement Uniqueness Loss (DUL).



    \item \textbf{w/o Tag Align:} This variant removes the contrastive tag alignment loss ($\mathcal{L}_{\text{align}}$) but still benefits from the direct classification signal of tag prediction and the uniqueness constraint.

    \item \textbf{w/o Tag Pred:} In this setup, we ablate the tag prediction loss ($\mathcal{L}_{\text{pred}}$). The model must rely solely on the contrastive alignment loss to structure its hierarchical latent space.

    \item \textbf{w/o DUL:} This variant is trained without the uniqueness loss ($\mathcal{L}_{\text{unique}}$), making it susceptible to ID collisions. To handle such collisions and ensure unique identifiers for the downstream recommender, we adopt the resolution strategy from TIGER~\cite{rajput2023recommender}, which appends an additional, sequentially incrementing integer to the end of any colliding semantic ID.
\end{itemize}

\begin{table}[h!]
\centering
\vspace{-5pt}
\caption{Ablation study results on Beauty and KuaiRand, detailing the individual contributions of Tag Alignment, Tag Prediction, and the Disentanglement via Uniqueness Loss.}
\vspace{-2mm}
\label{tab:ablation}
{%
\begin{tabular}{l|cc|cc}
\toprule
\multirow{2}{*}{\textbf{Methods}} & \multicolumn{2}{c|}{\textbf{Beauty}} & \multicolumn{2}{c}{\textbf{KuaiRand}} \\
 & \textbf{R@10} & \textbf{N@10} & \textbf{R@10} & \textbf{N@10} \\
\midrule
\hline
HiD-VAE & \textbf{0.0698} & \textbf{0.0421} & \textbf{0.0785} & \textbf{0.0586} \\
\midrule
w/o Tag Align & 0.0651 & 0.0392 & 0.0742 & 0.0541 \\
w/o Tag Pred  & 0.0633 & 0.0378 & 0.0725 & 0.0529 \\
w/o DUL       & 0.0524 & 0.0301 & 0.0668 & 0.0483 \\
\bottomrule
\end{tabular}%
}
\end{table}

The results of our ablation study are presented in Table~\ref{tab:ablation}. Our analysis yields the following key findings: 
\begin{itemize}[leftmargin=*]
    \item \textbf{Hierarchical Supervision is crucial for semantic grounding.} Removing either component of the HS mechanism leads to a noticeable drop in performance. Specifically, ablating the Tag Prediction loss (`w/o Tag Pred') results in a more significant decline than removing the Tag Alignment loss (`w/o Tag Align'). This suggests that while both components are vital, the direct classification signal from $\mathcal{L}_{\text{pred}}$ serves as a stronger semantic anchor, forcing each code to map to a concrete category. The contrastive $\mathcal{L}_{\text{align}}$, in turn, is essential for refining the geometric structure of the latent space, ensuring that semantically similar categories are represented closely. The synergy of both losses is key to learning a robust and meaningful hierarchy. For detailed architecture of the layer-specific tag predictors and their per-layer classification accuracies (which remain high even for fine-grained layers with numerous categories), see Appendix~\ref{tag_predictor}.

    \item \textbf{Disentanglement is paramount; avoiding ID collisions is important.} The most substantial performance degradation occurs in the `w/o DUL' variant. This finding is critical. While adopting the TIGER-style post-hoc fix (appending an integer) technically resolves ID collisions and prevents evaluation errors, it does so at a great semantic cost. This strategy injects non-semantic, arbitrary information into the final layer of the item identifier. For example, an ID that should purely represent a ``face mask'' might become `[beauty, skincare, mask, \textbf{1}]'. This appended integer disrupts the learned semantic sequence, effectively introducing noise that confuses the downstream generative model and hampers its ability to reason over the item's true attributes. This result strongly validates our core motivation: achieving intrinsic disentanglement during the representation learning stage is far superior to relying on superficial, post-processing fixes that corrupt the semantic integrity of the learned identifiers.
\end{itemize}

\subsection{Qualitative Insights and Analysis (RQ3)}\label{sec:qualitative_insights}

To gain deeper insights into the properties of the learned identifiers, we first provide a quantitative analysis of disentanglement, followed by a qualitative exploration of the learned hierarchical semantics.

\noindent\textbf{Effectiveness of Disentanglement.} A primary challenge for codebook-based identifiers is \textit{ID collision}, where multiple distinct items are mapped to the same discrete ID sequence. To rigorously evaluate our model's ability to mitigate this issue, we calculate the ID collision rate by determining the percentage of items that share non-unique IDs out of the total number of items. A lower rate is highly desirable, as it signifies a more robust one-to-one mapping between items and their learned identifiers, which is critical for recommendation accuracy and evaluation integrity.

\begin{table}[htbp]
\centering
\vspace{-1mm}
\caption{ID Collision Rate (\%) comparison on three datasets. Our full model, HiD-VAE, drastically reduces collisions to a negligible level. Lower values are better.}
\label{tab:collision}
\vspace{-2mm}
\begin{tabular}{lccc}
\toprule
\textbf{Methods} & \textbf{Beauty} & \textbf{Sports} & \textbf{KuaiRand} \\
\midrule
\quad VQ-Rec & 21.2\% & 22.5\% & 20.3\% \\
\quad RQ-VAE & 18.7\% & 19.5\% & 17.8\% \\
\midrule
\quad HiD-VAE (Full) & \textbf{2.1\%} & \textbf{2.8\%} & \textbf{1.9\%} \\
\quad \quad w/o DUL & 17.5\% & 18.2\% & 16.9\% \\
\quad \quad w/o HS  & 5.8\% & 6.5\% & 5.2\% \\
\bottomrule
\end{tabular}
\vspace{-1mm}
\end{table}

Table~\ref{tab:collision} presents a comparison of collision rates. We include two unsupervised baselines: `VQ-Rec', which uses product quantization, and `RQ-VAE', which uses residual quantization and forms the basis for models like TIGER and LC-REC. Our analysis reveals: 
\begin{itemize}[leftmargin=*]
    \item \textbf{Unsupervised tokenization methods inherently suffer from high collision rates.} Both baseline methods exhibit substantial ID collisions, with VQ-Rec's flat quantization scheme performing the worst (up to 22.5\% on Sports). This highlights a fundamental weakness in existing unsupervised approaches, where the lack of explicit constraints leads to significant representation entanglement. In stark contrast, our full HiD-VAE model reduces the collision rate to a negligible level (e.g., 2.8\% on Sports), representing a remarkable \textbf{87.6\% relative reduction} compared to the strongest baseline (VQ-Rec). This provides direct quantitative proof that our framework achieves a nearly injective mapping.

    \item \textbf{The Disentanglement Uniqueness Loss (DUL) is the primary driver of this success.} The `w/o DUL' variant, which lacks the $\mathcal{L}_{\text{unique}}$ objective, performs only marginally better than RQ-VAE. This unequivocally demonstrates that an explicit disentanglement mechanism is essential to prevent the representation collapse that plagues standard VQ-based tokenization schemes.

    \item \textbf{Hierarchical Supervision (HS) indirectly contributes to disentanglement.} An interesting finding is that the `w/o HS' variant, while significantly better than the baselines due to the presence of DUL, still incurs a higher collision rate than the full HiD-VAE model. This suggests that by imposing a strong, semantically meaningful structure on the latent space via tag alignment and prediction, our hierarchical supervision losses naturally encourage a better separation of item representations. This pre-structured space then allows DUL to operate more effectively, further enhancing the uniqueness of the final identifiers.
\end{itemize}

In summary, the results confirm that HiD-VAE, through the synergistic combination of disentanglement and structured semantic supervision, effectively resolves the critical ID collision problem inherent in  VQ- and RQ-based unsupervised tokenization methods.

\noindent\textbf{Visualization of the Disentanglement Effect.} To provide direct, qualitative evidence of the uniqueness loss's efficacy, we visualize the initial latent space ($\boldsymbol{z}_0$) of items prone to collision using t-SNE~\cite{maaten2008visualizing}. 
Using our `w/o DUL' ablation model on the Beauty dataset, we first identify the top 11 fine-grained categories with the highest ID collision rates. From each of these high-collision categories, we then select up to 50 distinct items that were erroneously mapped to shared identifiers. For these selected items, we extract their initial latent representations—the continuous vectors $\boldsymbol{z}_0$ produced by the encoder before quantization—from both the `w/o DUL' model and our full HiD-VAE model, which are then projected into a 2D space using t-SNE for visual comparison.

\begin{figure}[htbp]
    \centering
    \vspace{-5pt}
    \begin{subfigure}[b]{0.49\columnwidth}
        \centering
        \includegraphics[width=\textwidth]{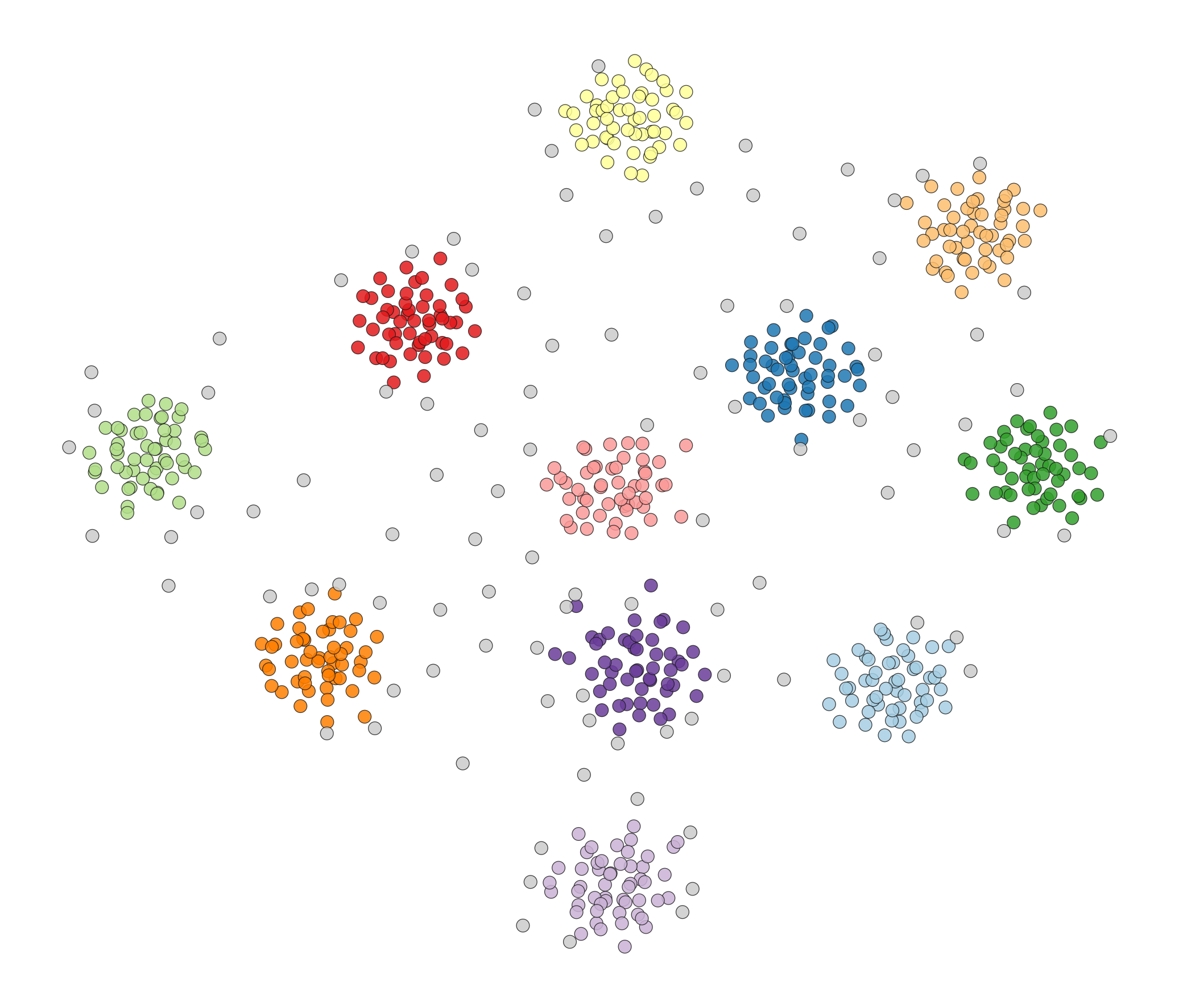}
        \caption{Without DUL (Entangled)}
        \label{fig:disentangle_a}
    \end{subfigure}
    \hfill
    \begin{subfigure}[b]{0.49\columnwidth}
        \centering
        \includegraphics[width=\textwidth]{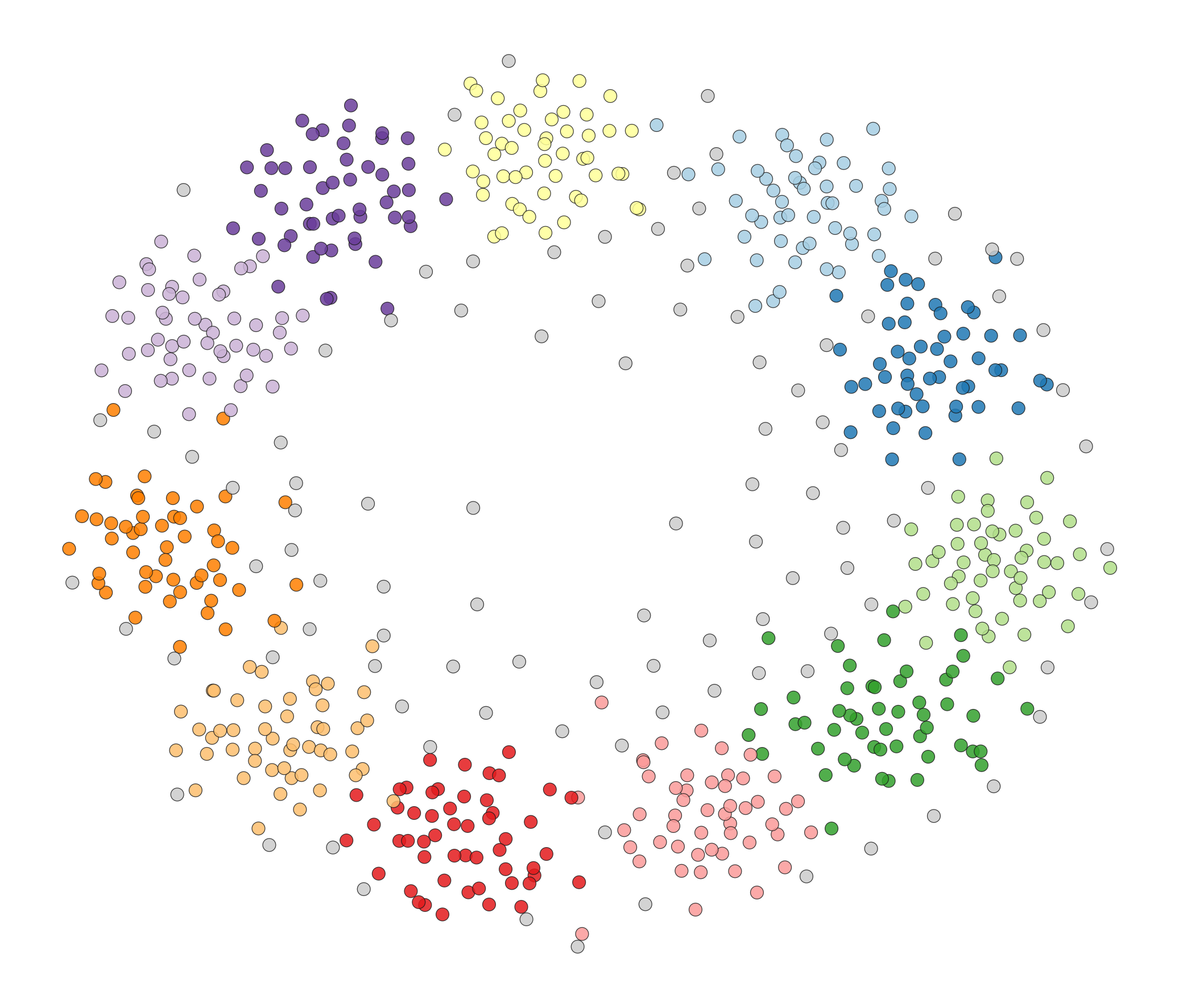}
        \caption{With DUL (Disentangled)}
        \label{fig:disentangle_b}
    \end{subfigure}
    \vspace{-2mm}
    \caption{The t-SNE visualization of the disentanglement effect of our Uniqueness Loss (DUL). Each color represents a group of distinct items from one of the top-11 high-collision categories. The gray dots represent items from other random categories, serving as a background to illustrate the overall structure of the latent space.}
    \label{fig:tsne_disentanglement}
    \vspace{-1pt}
\end{figure}

The results, presented in Figure~\ref{fig:tsne_disentanglement}, offer a stark visual contrast. Figure~\ref{fig:tsne_disentanglement}(a) displays the entangled latent space from the `w/o DUL' model, where items from the same group (represented by a single color) considerably collapse into overlapping representations. This tight clustering makes items indistinguishable and visually demonstrates the root cause of ID collisions. In contrast, Figure~\ref{fig:tsne_disentanglement}(b) shows the latent space from our full HiD-VAE model. Here, the structure is remarkably well-separated, as the DUL has effectively pushed the items within each group apart, ensuring they acquire unique representations. While items from the same category still occupy a similar semantic neighborhood, they are now clearly separable. This visually confirms that our uniqueness loss is highly effective at resolving representation entanglement at its source, creating a robust foundation for learning unique and meaningful identifiers.


\begin{figure}[!t]
    \centering
    \includegraphics[width=0.9\columnwidth]{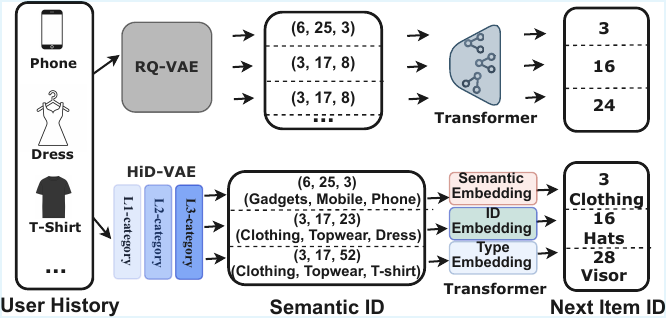}
    \vspace{-4pt}
    \caption{Case study comparing a standard RQ-VAE with our HiD-VAE. For the same user, HiD-VAE learns a transparent semantic path (e.g., `Skincare -> Treatments -> Serums'), enabling traceable reasoning. In contrast, the baseline's opaque codes (e.g., `[17, 83, 152]') result in black-box logic and risk generating invalid recommendations.}
    \label{fig:case_study}
    \vspace{-3mm}
\end{figure}

\noindent\textbf{Case Study on Recommendation Interpretability.}
To demonstrate the practical advantages of our framework beyond aggregate metrics, we conduct a case study comparing HiD-VAE with a standard RQ-VAE baseline. As shown in Figure~\ref{fig:case_study}, we analyze their behavior for a user from the Beauty dataset. RQ-VAE maps an item, like a serum, to an opaque identifier (e.g., `[17, 83, 152]'). The semantic meaning of these codes is unknown, rendering the generation process a black box that relies on statistical pattern matching. In contrast, HiD-VAE generates a transparent, self-explanatory ID (e.g., `[5, 12, 47]') that directly decodes into a human-readable path: `Skincare -> Treatments -> Serums'. This allows the generation process to follow a traceable reasoning. Furthermore, in the Transformer-based recommender, hierarchy-aware semantic embeddings are employed, where each token in the semantic ID is mapped to semantic vectors from an embedding model, concatenated with learnable ID embeddings and level-specific type embeddings. These embeddings are jointly learned, enriching the representation with explicit semantic information to better capture the coarse-to-fine semantic path, thereby enhancing the overall performance.

This fundamental representation difference directly impacts the final recommendations' quality and reliability. While both models might suggest a relevant item, the baseline can only offer a list of numeral IDs. More critically, its unconstrained generation can ``hallucinate'' invalid ID combinations that do not correspond to any real item. HiD-VAE, however, provides an explanation rooted in its explicit semantic path, enhancing user trust. Furthermore, because its generation is constrained by the learned tag hierarchy, it is mechanically robust against producing invalid IDs. This superior controllability and trustworthiness are crucial for deploying generative recommenders in real-world systems.


\section{Conclusion}
\label{sec:conclusions}

In this work, we introduced HiD-VAE, a novel framework for generative recommendation that addresses the core limitations of existing semantic ID tokenization: semantic flatness and representation entanglement. By pioneering a hierarchically-supervised quantization process and an uniqueness loss, HiD-VAE learns interpretable, disentangled item representations that align with multi-level tags and minimize ID collisions. Extensive experiments on three benchmarks demonstrate that HiD-VAE achieves state-of-the-art performance, significantly outperforming prior discriminative and generative models. Our approach not only unifies the recommendation pipeline but also enhances traceability, paving the way for more robust real-world deployments. For future directions, we plan to incorporate multi-modal data for richer representations and integrate LLMs for better sequential modeling. More detailed discussions on future directions are provided in Appendix~\ref{future}.

\balance
\bibliographystyle{ACM-Reference-Format}
\bibliography{9bibfile}

\appendix
\section{Related Work}
\label{sec:appendix_related}

Modeling the dynamic nature of user preferences is a central challenge in recommender systems \cite{wang2019sequential,chen2021modeling}. Early deep learning approaches, such as GRU4Rec \cite{hidasi2015session}, pioneered the use of recurrent neural networks~\cite{medsker2001recurrent} to capture sequential patterns in user interaction histories. More recently, the field has been dominated by Transformer-based architectures~\cite{vaswani2017attention}, which have proven more effective at modeling complex dependencies. Models like SASRec~\cite{kang2018self} adopted a decoder-only, self-attentive structure for next-item prediction, while BERT4Rec~\cite{sun2019bert4rec} introduced a bidirectional training objective inspired by masked language modeling in NLP. Subsequent work, such as S$^3$-Rec~\cite{zhou2020s3}, further refined these models by incorporating sophisticated self-supervised pre-training tasks to enhance representation quality. However, these methods are fundamentally discriminative; they learn item embeddings and then rely on an external Approximate Nearest Neighbor (ANN) search index, such as Faiss~\cite{Faiss}, to retrieve candidates from a massive item corpus by scoring their relevance. This separation of representation learning and retrieval has motivated a new paradigm.

A recent paradigm shift towards generative retrieval reframes recommendation as an autoregressive sequence generation task, where the model directly generates item identifiers instead of scoring candidates~\cite{rajput2023recommender,zheng2024adapting,wang2024learnable}. Pioneering this approach, TIGER~\cite{rajput2023recommender} introduced ``Semantic IDs''—discrete item representations learned via Residual-Quantized VAEs (RQ-VAE)~\cite{zeghidour2022soundstream} from item content features. Subsequent research, including LC-Rec~\cite{zheng2024adapting} and LETTER~\cite{wang2024learnable}, has focused on enhancing these IDs by integrating collaborative signals into the tokenization process. Despite their promise, these generative models face two critical limitations. First, the semantic space they learn is ``flat'' and uninterpretable \cite{singh2024better}; the hierarchy is an implicit byproduct of the quantization process rather than an explicitly supervised structure, rendering the models opaque \cite{zhu2020s3vae}. Second, they suffer from representation entanglement \cite{flam2022learning}, which leads to ``ID collisions''~\cite{rajput2023recommender,wu2025graphhash} where distinct items are mapped to the same identifier, harming recommendation diversity and accuracy. Existing solutions to this problem are often post-hoc and fail to address the root cause of entanglement in the latent space \cite{rajput2023recommender}. In contrast, our work directly tackles these dual challenges by introducing a hierarchically-supervised quantization process for interpretability and a uniqueness loss to enforce representation disentanglement at its source.

\section{Details on Hierarchical Tag Generation}
\label{prompt}

To extend HiD-VAE to datasets without native hierarchical tags, such as KuaiRand, we employ a structured LLM-based pipeline. Here, we provide specifics on the LLM selection and prompt design to ensure reproducibility and clarity.

We utilize Qwen3-235B-A22B-Instruct-2507~\cite{bai2023qwen}, a state-of-the-art large language model known for its strong reasoning and classification capabilities in constrained tasks. This model was selected for its efficiency in handling multi-turn prompts and its robustness against hallucinations when provided with candidate sets, aligning well with our retrieval-then-classification paradigm.

The prompt templates are designed to be hierarchical and context-aware, incorporating prior level tags to enforce consistency across the category path. We use a two-template system: one for initial candidate retrieval (implicit via embeddings) and another for LLM classification. The classification prompt is structured to include the item's textual content, previous tags, and candidates, phrased as a zero-shot classification task to minimize bias.

For illustration, Figure~\ref{fig:llm_prompt} shows an abridged example of the LLM classification prompt template, applied sequentially for each level $l$. In practice, full prompts include additional instructions for edge cases (e.g., ambiguous content) and are formatted in JSON for structured output parsing.

Empirical tuning revealed that including 5--10 candidates per level ($K=10$ in experiments) balances precision and computational cost, with the LLM achieving over 95\% intra-batch consistency in tag hierarchies on validation subsets. This pre-processing step is performed offline once per dataset.

\begin{figure}[htbp]
\lstset{style=custom_style}
\centering
\begin{lstlisting}%[basicstyle=\small\ttfamily, frame=single, breaklines=true, breakindent=0pt, captionpos=b]
You are a category expert specializing in hierarchical classification. 
Given the item description: "{item_text}", and the previous hierarchical tags: {prev_tags}, please select the single best matching tag for the next level from the following candidate list: {candidates}. 
Reason step-by-step if needed, but output only the selected tag in plain text, without any additional explanation or formatting.
\end{lstlisting}
\caption{Abridged LLM classification prompt template for zero-shot hierarchical tag selection. }
\label{fig:llm_prompt}
\end{figure}

\section{Details on HiD-VAE Algorithm}
\subsection{Autoencoder Objective}
\label{autoencoder}

The reconstruction loss ($\mathcal{L}_{\text{recon}}$) ensures the model can faithfully reconstruct the original input $\boldsymbol{x}$. We use the Mean Squared Error (MSE) between the input and its reconstruction $\hat{\boldsymbol{x}}$:
\begin{equation}
    \mathcal{L}_{\text{recon}} = ||\boldsymbol{x} - \hat{\boldsymbol{x}}||_2^2
\end{equation}

Furthermore, to train the discrete bottleneck, the commitment loss ($\mathcal{L}_{\text{commit}}$) regularizes the encoder. It encourages the encoder's continuous output $\boldsymbol{z}_e(\boldsymbol{x})$ to commit to its chosen codeword $\boldsymbol{z}_q(\boldsymbol{x})$. This is managed by a stop-gradient ($\text{sg}$) operator that isolates the gradient flow to only update the encoder for this term~\cite{van2017neural}. The final loss is summed over all $L$ quantization stages:
\begin{equation}
    \mathcal{L}_{\text{commit}} = ||\boldsymbol{z}_e(\boldsymbol{x}) - \text{sg}[\boldsymbol{z}_q(\boldsymbol{x})]||_2^2
\end{equation}

\subsection{Algorithmic Overview and Implementation Details}
\label{algorithm}

In this section, we provide a concise algorithmic overview of the HiD-VAE framework, complementing the detailed descriptions in the main text. We focus on presenting a unified pseudocode for the end-to-end process, including hierarchical tag generation, representation learning, and sequential recommendation. This serves to clarify the integration of components without reiterating the core innovations (e.g., hierarchical supervision and uniqueness loss) already elaborated in Sections~\ref{sec:Hierarchical_Tag_Generation}--\ref{sec:Interpretable_Generative-Recommendation}. Additionally, we briefly discuss implementation considerations such as batch processing and inference efficiency, which enhance practical deployment.

The pseudocode in Algorithm~\ref{alg:hidvae} outlines the two-stage training and inference pipeline. For brevity, we abstract low-level operations like encoder/decoder forward passes and assume standard PyTorch-style implementations. Key hyperparameters (e.g., $\beta$ terms, $L$, $K_l$) are as defined in the main text; in practice, we tune them via grid search on validation sets, with typical values including $\tau=0.07$, $m=0.9$, and batch sizes of 512 for Stage 1.

\begin{algorithm}[bt!]
\caption{HiD-VAE Framework}
\label{alg:hidvae}
\begin{algorithmic}[1]
\Require Item features $\{\boldsymbol{x}_i\}_{i \in \mathcal{I}}$, optional tags $\{\{c_i^{(l)}\}_{l=1}^L\}_{i \in \mathcal{I}}$, user sequences $\{\mathcal{S}_u\}_{u \in \mathcal{U}}$
\Ensure Predicted next items for each user

\Statex \textit{// Stage 0: Hierarchical Tags Generation (if tags unavailable)}
\For{each item $i \in \mathcal{I}$}
    \State $\boldsymbol{v} \gets \mathcal{E}_{\text{sent}}(\boldsymbol{x}_{i,\text{text}})$  \Comment{Semantic embedding}
    \For{$l = 1$ to $L$}
        \State $\mathcal{C}_{\text{cand}}^{(l)} \gets$ Top-K retrieval from $\mathcal{T}^{(l)}$ via sim($\boldsymbol{v}$, tag embeds)
        \State $t^{(l)*}, c^{(l)*} \gets$ LLM classify($\boldsymbol{x}_{i,\text{text}}, \{t^{(j)}\}_{j< l}, \mathcal{C}_{\text{cand}}^{(l)}$)
    \EndFor
\EndFor

\Statex \textit{// Stage 1: HiD-VAE Training}
\State Initialize encoder $E$, decoder $D$, codebooks $\{\mathcal{C}^{(l)}\}_{l=1}^L$, projectors $\{P_l\}_{l=1}^L$, classifiers $\{C_l\}_{l=1}^L$
\While{not converged}
    \State Sample batch $\{\boldsymbol{x}_b, \{\boldsymbol{t}_b^{(l)}\}_{l=1}^L\}_{b=1}^B$
    \State $\boldsymbol{z}_0 \gets E(\boldsymbol{x})$  \Comment{Batch-wise}
    \State $\boldsymbol{r}_0 \gets \boldsymbol{z}_0$
    \For{$l = 1$ to $L$}
        \State $\boldsymbol{e}^{(l)}, y^{(l)} \gets q_l(\boldsymbol{r}_{l-1})$  \Comment{Quantize residual}
        \State $\boldsymbol{r}_l \gets \boldsymbol{r}_{l-1} - \boldsymbol{e}^{(l)}$
        \State $\boldsymbol{z}_q^{(l)} \gets \sum_{j=1}^l \boldsymbol{e}^{(j)}$
        \State Compute $\mathcal{L}_{\text{align}}^{(l)}$, $\mathcal{L}_{\text{pred}}^{(l)}$
    \EndFor
    \State $\hat{\boldsymbol{x}} \gets D(\boldsymbol{z}_q^{(L)})$
    \State Compute $\mathcal{L}_{\text{recon}}$, $\mathcal{L}_{\text{commit}}$, $\mathcal{L}_{\text{unique}}$ (on colliding pairs)
    \State Update parameters via $\mathcal{L}_{\text{HiD-VAE}}$
\EndWhile
\State Freeze HiD-VAE; map all items to IDs $\{\boldsymbol{y}_i\}_{i \in \mathcal{I}}$

\Statex \textit{// Stage 2: Recommender Training}
\State Initialize Transformer with hierarchy-aware embeddings
\While{not converged}
    \State Sample user batch $\{\mathcal{S}_u\}$
    \State Map to ID sequences $\{(\boldsymbol{y}_1, \ldots, \boldsymbol{y}_{T})\}$
    \State Enrich embeddings with tag semantics and level types
    \State Compute $\mathcal{L}_{\text{rec}}$ via autoregressive CE
    \State Update Transformer parameters
\EndWhile

\Statex \textit{// Inference}
\For{each test sequence $(\boldsymbol{y}_1, \ldots, \boldsymbol{y}_T)$}
    \State Autoregressively generate $\boldsymbol{y}_{T+1}$ with constrained decoding (mask invalid prefixes)
    \State Map to item via pre-computed ID-to-item index
\EndFor

\end{algorithmic}
\end{algorithm}

During inference, the constrained decoding in Stage 2 utilizes a trie-based prefix tree for efficient vocabulary masking, reducing invalid generations to zero while maintaining $O(L \log K)$ time per token (where $K = \max K_l$). This structure is built offline from all valid item IDs, ensuring scalability for large catalogs ($|\mathcal{I}| \sim 10^6$).

We also note that for datasets with partial tags, a hybrid approach can interpolate LLM-generated and ground-truth labels, weighted by confidence scores from the LLM. Empirical ablation shows this boosts alignment loss convergence by 15--20\% in early epochs, though full results are omitted for space.

\section{Details on Datasets and Baselines}

\subsection{Datasets}\label{app:dataset_stats}
\begin{table}[htbp]
    \centering
    \caption{Statistics of the experimental datasets.}
    \label{tab:dataset_statistics}
    \begin{tabular}{lrrrr}
        \toprule
        \textbf{Dataset} & \textbf{\#Users} & \textbf{\#Items} & \textbf{\#Interactions} & \textbf{\#Seq.Length} \\
        \midrule
        Beauty              & 22,363 & 12,101 & 198,360 & 8.87 \\
        Sports & 35,598 & 18,357 & 296,175 & 8.32 \\
        KuaiRand      & 983 & 29,983 & 953,166 & 19.83 \\
        \bottomrule
    \end{tabular}
\end{table}

\subsection{Baselines}
\label{baselines}

\begin{itemize}[leftmargin=*]
    \item \textbf{Traditional Sequential Models:} This group includes classic and influential non-Transformer models that serve as strong foundational baselines.
    \textbf{GRU4Rec}~\cite{hidasi2015session} is a pioneering model that applies Gated Recurrent Units (GRUs) to model user sessions.
    \textbf{Caser}~\cite{tang2018personalized} employs Convolutional Neural Networks (CNNs) to capture local sequential patterns as ``images'' over the item embedding matrix.
    \textbf{HGN}~\cite{ma2019hierarchical} utilizes a hierarchical gating network to integrate long-term and short-term user interests.
    \textbf{NextItNet}~\cite{yuan2019simple} leverages a stack of dilated convolutional layers to efficiently capture long-range dependencies in user sequences.

    \item \textbf{Transformer-based Models:} These models represent the current standard for discriminative sequential recommendation, using self-attention to capture complex user dynamics.
    \textbf{SASRec}~\cite{kang2018self} first applied the Transformer's self-attention mechanism to sequential recommendation. \textbf{BERT4Rec}~\cite{sun2019bert4rec} adapts the deep bidirectional pre-training paradigm of BERT, using a cloze task to learn user behavior representations.


    \item \textbf{Generative Recommendation Models:} This category includes several key generative baselines. \textbf{TIGER}~\cite{rajput2023recommender} employs a hierarchical RQ-VAE to learn semantic item identifiers, which are subsequently predicted by a Transformer-based sequence model. Similarly, \textbf{LC-Rec}~\cite{zheng2024adapting} also leverages RQ-VAE based identifiers but focuses on aligning language and collaborative semantics. To provide a non-hierarchical contrast, we adapt \textbf{VQ-Rec}~\cite{hou2023learning} to a generative framework by training a Transformer on its semantic IDs generated via Product Quantization.

\end{itemize}

\section{Details on Experiments}
\label{experiments}

\subsection{Tag Predictor Architecture and Per-Layer Accuracy}
\label{tag_predictor}

The tag predictor for each layer is a multi-layer perceptron with enhancements for robustness: it begins with a self-attention mechanism to weigh input features, followed by a feature extraction layer (linear projection to a hidden dimension, LayerNorm if enabled, ReLU activation, and dropout), two residual blocks (each consisting of linear projections with intermediate LayerNorm, ReLU, and dropout for feature refinement), and a final classification head (multiple linear layers with LayerNorm, ReLU, and reduced dropout). Key hyperparameters adapt per layer: the hidden dimension starts at twice the input embedding size and scales up (e.g., multiplied by $(l + 1)$), while dropout increases gradually to prevent overfitting on complex, fine-grained predictions. Batch normalization is optionally applied throughout for stable training.

\begin{table}[htbp]
\centering
\caption{Per-layer tag prediction accuracy (\%) on Beauty, Sports, and KuaiRand datasets.}
\vspace{-1mm}
\label{tab:tag_accuracy}
\setlength{\tabcolsep}{1.3mm}
\begin{tabular}{lccc}
\toprule
Dataset & Layer 1 (Coarse) & Layer 2 (Medium) & Layer 3 (Fine) \\
\midrule
Beauty & 96.38(7) & 92.73(30) & 85.61(97) \\
Sports & 93.68(23) & 87.54(70) & 83.27(119) \\
KuaiRand & 87.84(38) & 83.58(97) & 77.49(146) \\
\bottomrule
\end{tabular}
\end{table}

To evaluate the effectiveness of these layer-specific predictors, we report per-layer classification accuracies on three datasets in Table~\ref{tab:tag_accuracy}, where the numbers in parentheses denote the effective number of tag categories at each level (after filtering out extremely rare categories with fewer than 30 samples to mitigate class imbalance and focus on well-represented semantics). The results indicate consistently strong performance across layers, with accuracies ranging from 96.38\% at the coarsest level (Layer 1) to as high as 85.61\% at the finest level (Layer 3) on the Beauty dataset, despite the progressive increase in category granularity and count. Notably, even as the number of categories escalates to 97--146 in deeper layers—yielding multi-class problems where random chance accuracy would be as low as approximately 0.68\%--1.03\%—the achieved accuracies remain robust and substantially outperform naive baselines, underscoring the efficacy of our scaled classifier architecture in handling complex, hierarchical semantic distinctions. This level of performance is particularly satisfactory given the inherent challenges of fine-grained classification with diverse and numerous categories, validating the framework's ability to capture nuanced item representations without significant degradation in predictive reliability.


\subsection{Hyperparameter Sensitivity}
\label{hyperparam_sensitivity}

\noindent\textbf{Sensitivity to Number of Quantization Layers in HiD-VAE.} To further assess the robustness of our HiD-VAE framework, we conduct a sensitivity analysis on the number of quantization layers $L$, a critical hyperparameter that determines the depth of the hierarchical semantic IDs. We evaluate performance for $L \in \{3, 4, 5\}$, focusing on Recall@10 and NDCG@10 metrics across the Beauty and KuaiRand datasets. Note that we exclude results for $L=2$, as preliminary experiments revealed that two-layer semantic IDs suffer from severe representation limitations, including high ID collision rates and insufficient semantic granularity to capture fine-grained item distinctions, rendering the learned representations inadequate for effective generative recommendation.


\begin{figure}[htbp]
    \centering
    \begin{subfigure}[b]{0.48\columnwidth}
        \centering
        \includegraphics[width=\textwidth]{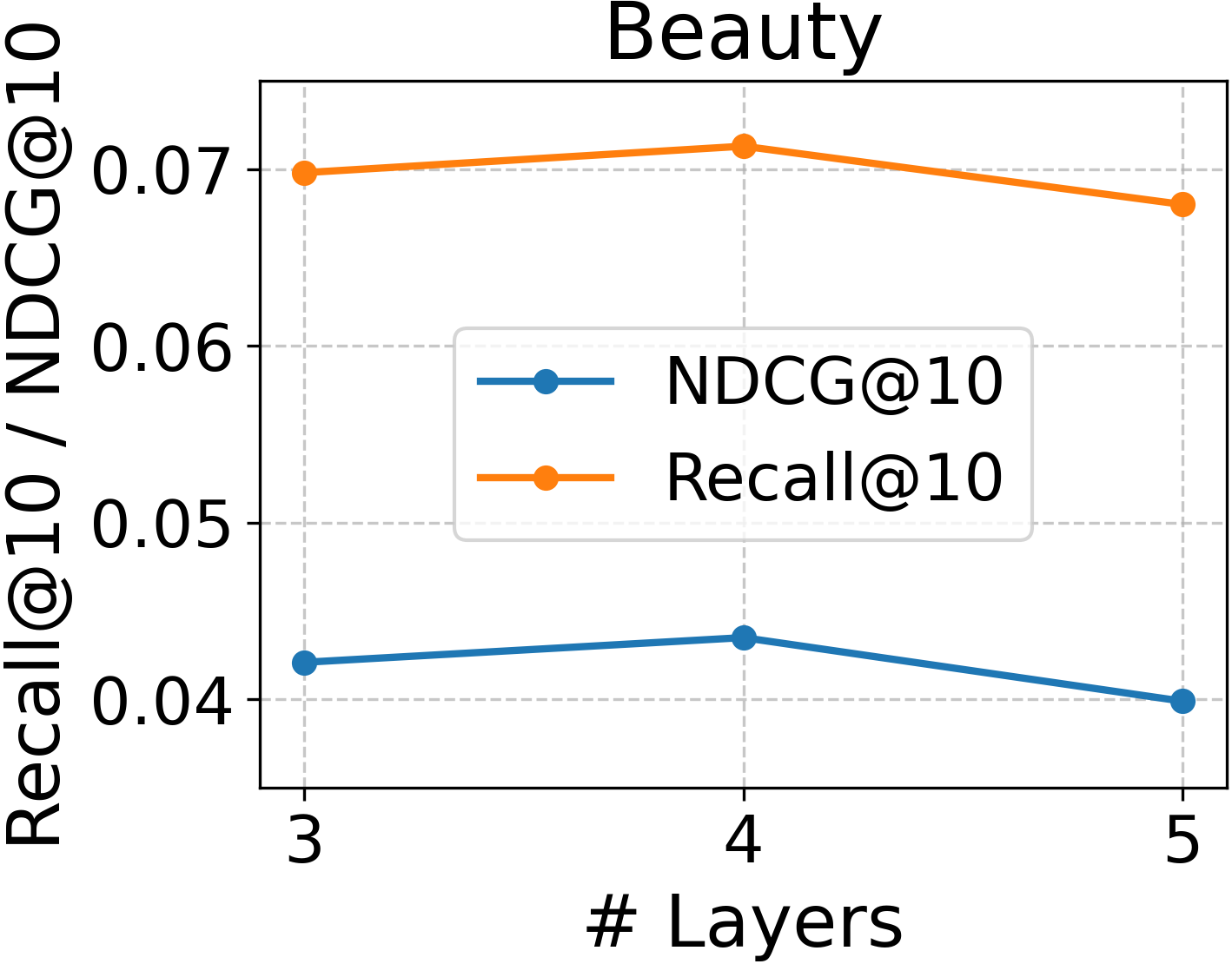}
        \caption{Beauty}
        \label{fig:layer_sensitivity_beauty}
    \end{subfigure}
    \hfill
    \begin{subfigure}[b]{0.48\columnwidth}
        \centering
        \includegraphics[width=\textwidth]{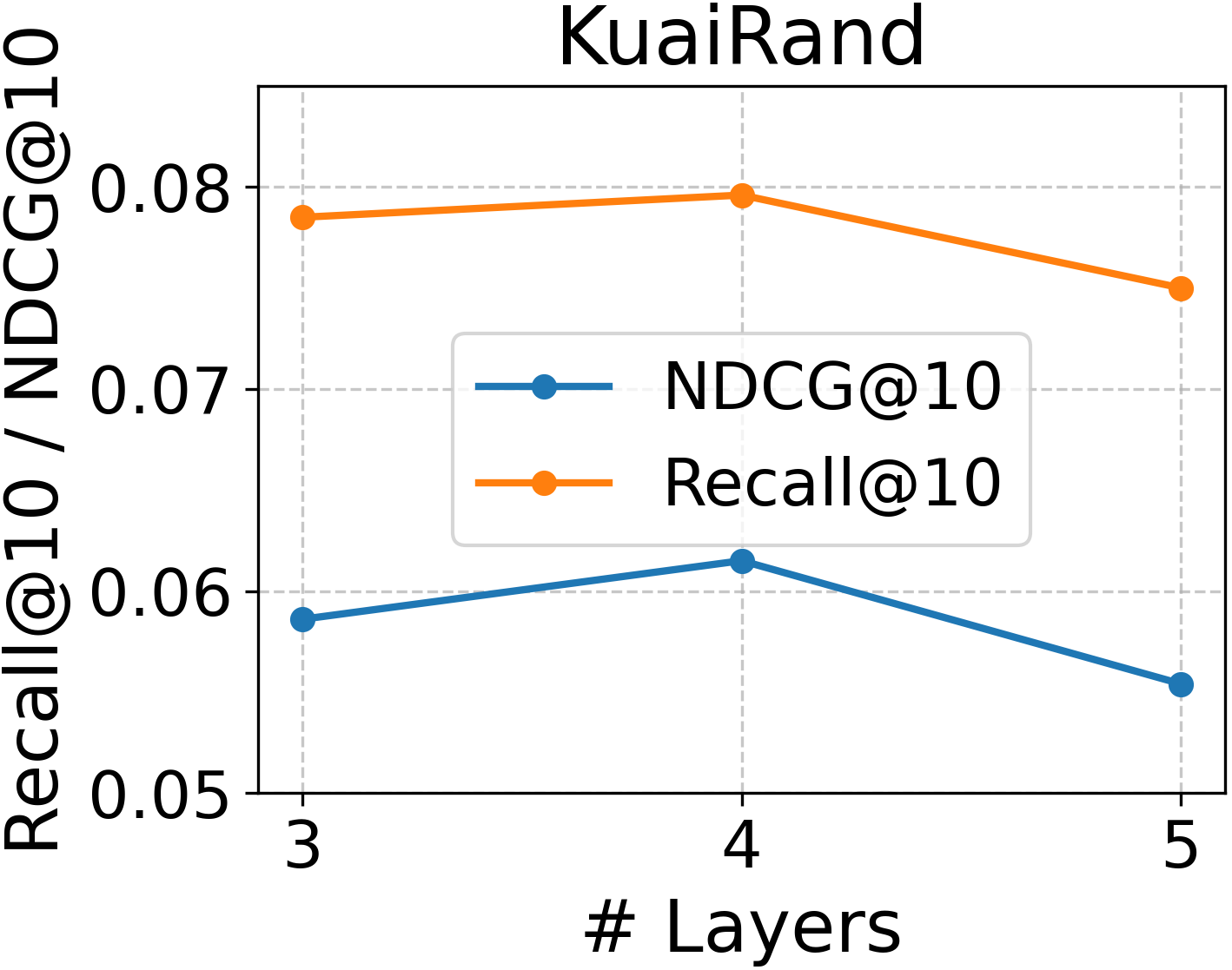}
        \caption{KuaiRand}
        \label{fig:layer_sensitivity_kuairand}
    \end{subfigure}
    \vspace{-3mm}
    \caption{Sensitivity analysis of the number of quantization layers $L$ on Recall@10 and NDCG@10. Performance peaks at $L=4$ but diminishes at $L=5$, indicating diminishing returns with increased depth.}
    \label{fig:layer_sensitivity}
\end{figure}

The results are visualized in Figure~\ref{fig:layer_sensitivity}. We observe that increasing $L$ from 3 to 4 yields marginal improvements (e.g., +4.88\% in Recall@10 on Beauty and +4.97\% on KuaiRand), as an additional layer enables finer semantic decomposition. However, further increasing to $L=5$ leads to performance degradation (e.g., -5.86\% in Recall@10 on Beauty relative to $L=4$), even falling below the $L=3$ baseline in some cases. This diminishing returns phenomenon highlights a low benefit-to-cost ratio, with deeper hierarchies elevating training costs—due to increased quantization overhead in Stage 1 and longer sequences in Stage 2—without proportional gains, likely from optimization challenges like propagating quantization errors and harder autoregressive modeling.

These findings underscore the importance of balancing hierarchy depth with computational efficiency, validating our default choice of $L=3$ as an optimal trade-off that achieves strong performance without excessive overhead.

\begin{figure}[htbp]
    \centering
    \begin{subfigure}[b]{0.45\columnwidth}
        \centering
        \includegraphics[width=\textwidth]{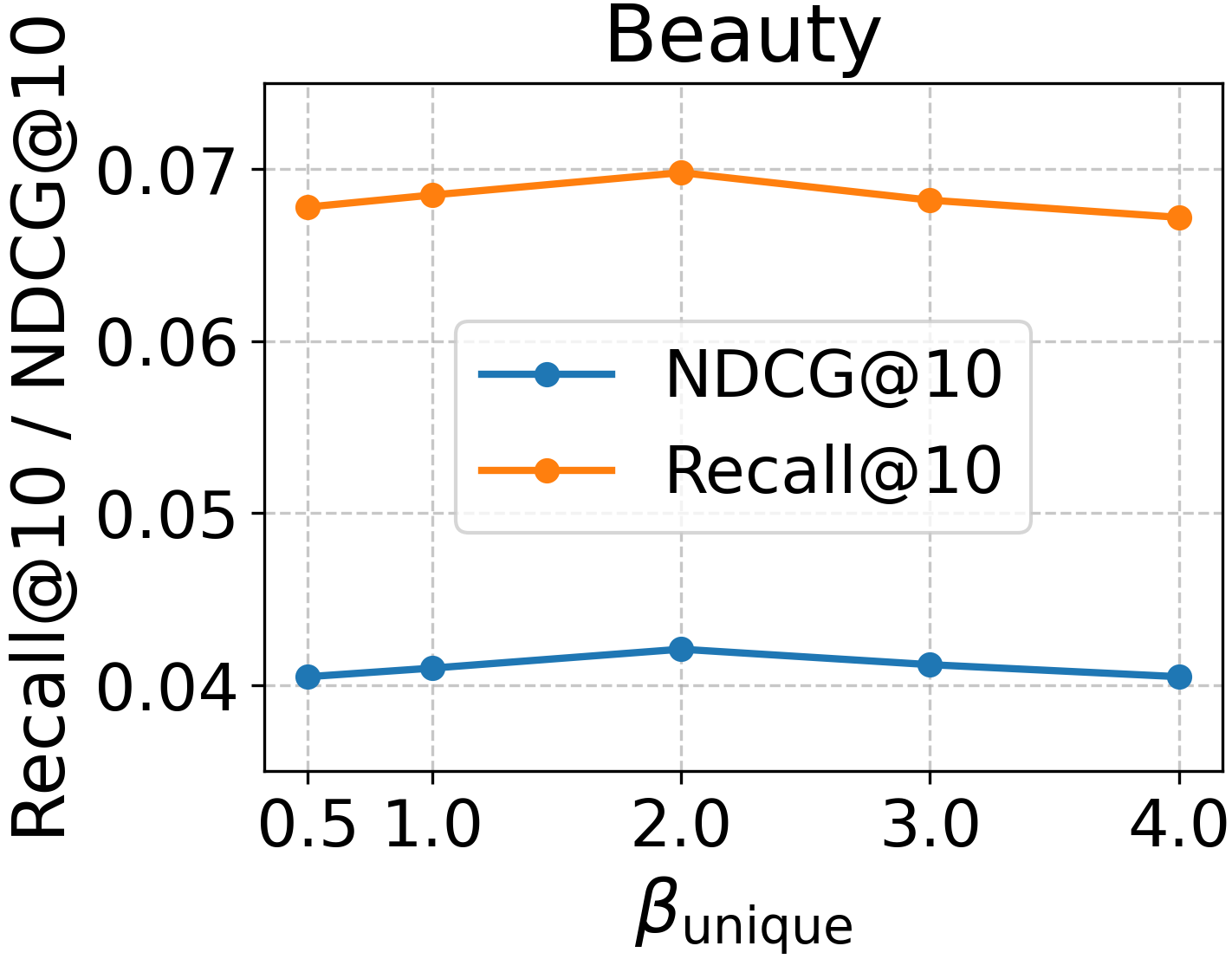}
        \caption{$\beta_{\text{unique}}$}
        \label{fig:beta_sensitivity}
    \end{subfigure}
    \hfill
    \begin{subfigure}[b]{0.45\columnwidth}
        \centering
        \includegraphics[width=\textwidth]{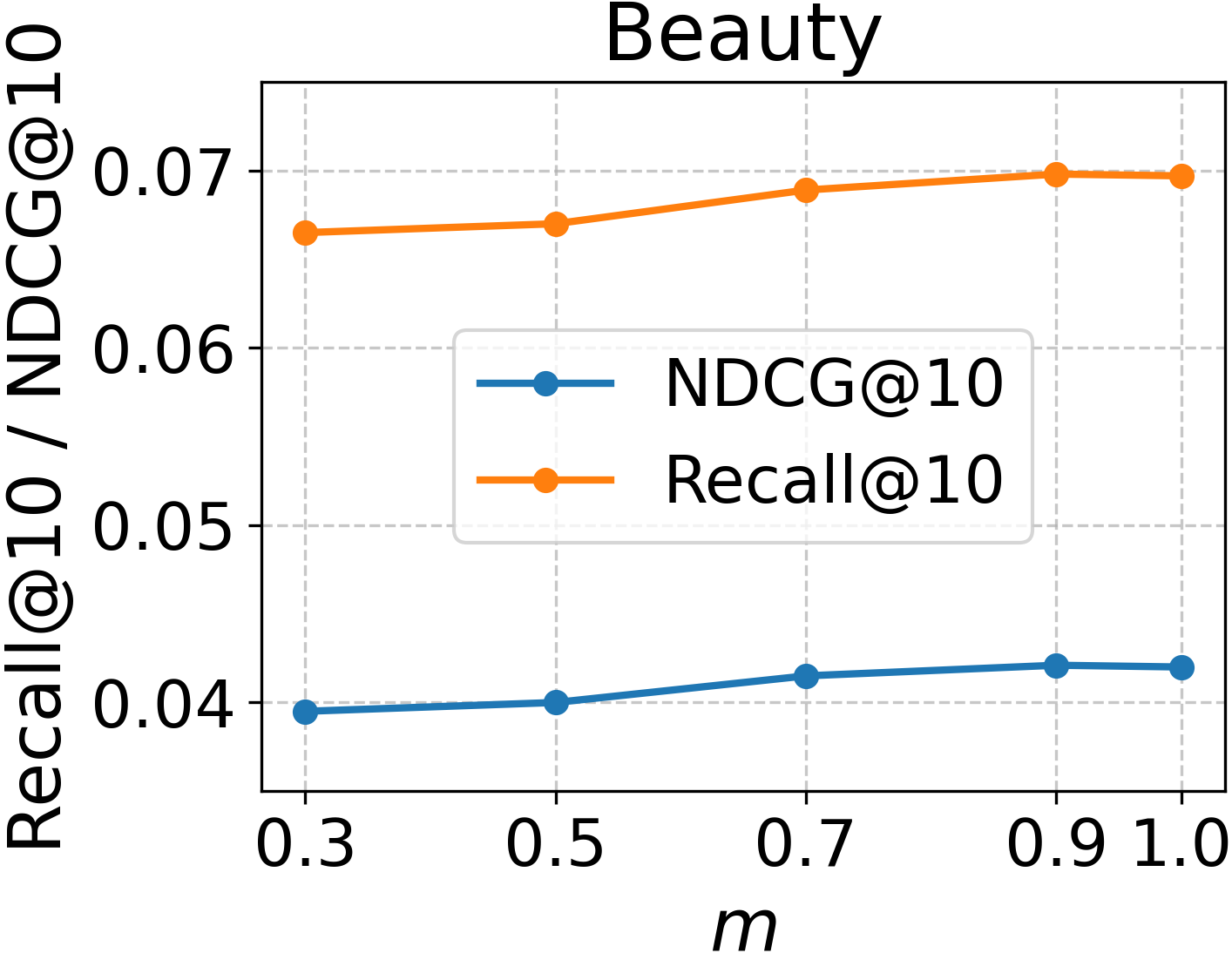}
        \caption{$m$}
        \label{fig:m_sensitivity}
    \end{subfigure}
    \vspace{-1mm}
    \caption{Sensitivity analysis of the uniqueness loss weight $\beta_{\text{unique}}$ and uniqueness margin $m$ on Recall@10 and NDCG@10 for the Beauty dataset. Performance peaks at $\beta_{\text{unique}}=2.0$ but diminishes at higher values, while metrics improve with increasing $m$ up to $0.9$, indicating optimal disentanglement with balanced penalties.}
    \label{fig:sensitivity}
\end{figure}

\noindent\textbf{Analyzing the uniqueness loss parameters.} To evaluate the robustness of HiD-VAE to the uniqueness loss weight $\beta_{\text{unique}}$, we vary it across \{0.5, 1.0, 2.0, 3.0, 4.0\} while keeping other parameters fixed. As shown in Figure~\ref{fig:beta_sensitivity} for the Beauty dataset, performance peaks at $\beta_{\text{unique}}=2.0$ with Recall@10=0.0698 and NDCG@10=0.0421, indicating an optimal balance where disentanglement is sufficiently enforced without overpowering other objectives. Lower values yield slightly degraded metrics due to increased ID collisions, while higher values may over-penalize and disrupt latent space structure.

We further analyze the uniqueness margin $m$ by testing values in \{0.3, 0.5, 0.7, 0.9, 1.0\}. Figure~\ref{fig:m_sensitivity} illustrates that on the Beauty dataset, metrics improve with increasing $m$, achieving the best results at $m=0.9$ (Recall@10=0.0698, NDCG@10=0.0421), as a stricter margin better separates colliding representations. Lower margins allow more overlap, reducing disentanglement efficacy.

\section{Future Directions}
\label{future}

For future directions, we plan to extend HiD-VAE's ID generation to multi-modal semantic IDs, integrating images, videos, and audio with text via encoders like CLIP. This would align discrete codes with cross-modal hierarchies, capturing nuances (e.g., visual styles in fashion, auditory patterns in music) for better accuracy and interpretability in heterogeneous domains like e-Commerce and social networks. In parallel, we aim to enhance the generative stage with large-scale language models (LLMs) for advanced sequential modeling. Fine-tuning or prompting LLMs on hierarchical IDs would leverage their reasoning for complex intents and dependencies, using hybrid architectures with rationales or counterfactuals to boost diversity and personalization. Broader efforts include diffusion models for ID synthesis, improving long-tail coverage and fairness, alongside privacy via federated learning.

\end{document}